\documentclass[twocolumn,trackchanges]{aastex63}
\usepackage{graphicx}
\usepackage{amssymb}
\usepackage{amsmath}
\usepackage{epstopdf}
\usepackage{calc}
\usepackage{multirow}
\newcommand{\T}{\ensuremath{\mathrm{T}}}

\shorttitle{SFRD evolution in the MeerKAT DEEP2 field}
\shortauthors{Matthews et al.}

\begin{document}

\pdfsuppresswarningpagegroup=1
\maxdeadcycles=1000

\title{
Confirmation of a Substantial Discrepancy between Radio and UV--IR Measures of the Star Formation Rate Density at $0.2 < z < 1.3$
}

\correspondingauthor{Allison M. Matthews} \email{amatthews@carnegiescience.edu}

\author[0000-0002-6479-6242]{A.~M.~Matthews}
  \affiliation{Observatories of the Carnegie Institution for Science,
  Pasadena, CA 91101, USA}

\author[0000-0003-4727-4327]{D.~D.~Kelson}
  \affiliation{Observatories of the Carnegie Institution for Science, 
  Pasadena, CA 91101, USA}

\author[0000-0001-7769-8660]{A.~B.~Newman}
  \affiliation{Observatories of the Carnegie Institution for Science, 
  Pasadena, CA 91101, USA}

\author[0000-0002-1873-3718]{F.~Camilo}
\affiliation{South African Radio Astronomy
  Observatory (SARAO), 2 Fir Street, Black River Park, Observatory,
  7925, South Africa}
  
\author[0000-0003-4724-1939]{J.~J.~Condon}
\affiliation{Unaffiliated}

\author[0000-0001-7363-6489]{W.~D.~Cotton}
  \affiliation{National Radio Astronomy
  Observatory, 520 Edgemont Road, Charlottesville, VA 22903, USA}

\author[0000-0001-5414-5131]{M.~Dickinson}
\affiliation{NSF's National Optical-Infrared Astronomy Research Laboratory, 
    950 N. Cherry Avenue, Tucson, AZ 85719, USA}

\author[0000-0002-4939-734X]{T.~H.~Jarrett}
\affiliation{Department of Astronomy, University of Cape Town, Rondebosch, 7700, South Africa}

\author[0000-0002-3032-1783]{M.~Lacy}
\affiliation{National Radio Astronomy
  Observatory, 520 Edgemont Road, Charlottesville, VA 22903, USA}

\begin{abstract}
We present the initial sample of redshifts for 3,839 galaxies in the MeerKAT DEEP2 field---the most sensitive $\sim$1.4\,GHz radio field yet observed with $\sigma_n=0.55\,\mu\,\mathrm{Jy\,beam^{-1}}$, reaching the confusion limit. Using a spectrophotometric technique combining coarse optical spectra with broadband photometry, we obtain redshifts with $\sigma_z \lesssim 0.01(1+z)$, as determined from repeat observations. The resulting radio luminosity functions between $0.2<z<1.3$ from our sample of 3,839 individual galaxies are in remarkable agreement with those inferred from previous modeling of radio source counts, confirming a $\gtrsim$ 50\% excess in radio-based SFRD$(z$) measurements at $0.2<z<1.3$ compared to those from the UV--IR. Several sources of systematic error are discussed---totalling $\sim$0.13\,dex when added in quadrature. Even in the event that all systematic errors work to decrease the radio-based SFRD values, they are incapable of reconciling differences between the radio-based measurements with those from the UV--IR at $0.5<z<1.3$. We conclude that significant work remains to have confidence in a full accounting of the star formation budget of the universe. 
\end{abstract}x

\keywords{galaxies: evolution -- galaxies: star formation -- galaxies:
  statistics -- radio continuum: galaxies}



\section{Introduction}
\nocite{matthews21a}
Since it was first discovered in the 1990s that the comoving star-formation rate density (SFRD, usually specified in $M_\odot\,\mathrm{yr^{-1}\,Mpc^{-3}}$) around $z\sim1$ dwarfs that of today, immense progress has been made in understanding the general shape of SFRD evolution with cosmic time. Collections of SFRD measurements across wide redshift ranges place the peak of star formation activity near $z\sim2$ and show a decline by a factor of $\sim$10 to present day \citep[see][for reviews of the topic]{hopkins06, madau14}. 



Despite consistent agreement among the UV/optical and infrared (IR) communities on the SFRD through $z\sim2$, there has been established tension between the SFRD evolution derived at shorter wavelengths (e.g. UV and optical) with that derived at longer wavelengths (e.g. radio and sub-millimeter). Studies with ALMA have shown that UV measurements of the SFRD miss the obscured, dusty, and heavily star-forming galaxies more common in the distant universe \citep[e.g.][]{casey18, bouwens20}. Even in the more ``recent'' past, \cite{whitaker17} found that $>80\%$ of star formation is obscured at all redshifts $z<2.5$ in galaxies having stellar masses $\log[M_*/M_\odot]\geq10$. FIR/sub-mm observations are essential to understand dusty galaxies, but single-dish telescopes are only sensitive to the massive, tip-of-the-iceberg sources. Sub-mm interferometers such as ALMA are limited to a very small field-of-view, making it expensive and impractical to amass a deep sample over a large enough area to minimize cosmic variance. 

Radio observations are immune to dust obscuration, insensitive to contamination by older stellar populations, and available across wide sky areas. While active galactic nuclei (AGNs) are primarily responsible for powering strong 1.4\,GHz radio sources, star-forming galaxies dominate the radio emission below $450\,\mu$Jy \citep[][]{algera20,matthews21b}. Star-forming galaxies produce radio emission through two processes: (1) thermal bremsstrahlung from \ion{H}{2} regions ionized and heated by massive stars and (2) synchrotron radiation from cosmic-ray electrons accelerated in the shocks of supernova remnants \citep{condon92}.

The tight relationship between the radio synchrotron and FIR luminosities of star-forming galaxies, described by the ``$q$'' parameter

\begin{equation}\label{eq:firrc}
    q \equiv \log\left[\frac{\mathrm{FIR}/(3.75\times10^{12}\mathrm{Hz}}{S_\mathrm{1.4\,GHz}\,(\mathrm{W\,m^{-2}\,Hz^{-1}})}\right],
\end{equation}
where
\begin{align}
    \mathrm{FIR}\,&\mathrm{(W\,m^{-2})} \equiv \qquad \\ &1.26\times10^{-14}[2.58\,S_\mathrm{60\,\mu m}\,\mathrm{(\mathrm{Jy})}+S_\mathrm{100\,\mu m} (\mathrm{Jy})]
\end{align}
\citep{helou88} was first derived in the local universe \citep{helou85}. It has been shown to evolve to varying degrees (from not-at-all to moderately) through redshift $z \sim 4$ \citep[e.g.][]{ivison10,magnelli15,pannella15,delhaize17}. More recently, \cite{delvecchio21} found that while the FIR/radio correlation is largely invariant with redshift, there is a statistically significant dependence on stellar mass $dq/d\log(M_*) = -0.148\pm0.013$. Given the established relationship between stellar mass and star formation rate \citep[e.g.][]{whitaker12}, this result is largely consistent with the nonlinear local FIR/radio correlation $dq/d\log(L_\nu)=-0.147$ of \cite{matthews21b}, derived from the largest, clean (free from AGN-dominated radio sources) sample of star-forming galaxies ($\sim4,300$) in the local universe. Further, previous studies have also found non-linearity in the FIR/radio correlation that may explain much of the perceived evolution in the $q$ parameter as a function of redshift \citep[e.g.][]{basu15,molnar21}.

Unfortunately, galaxies whose radio emission is powered primarily by star-formation are very faint. It is necessary to detect sub-$\mu$Jy sources to account for most of the star formation through $z\sim1-3$ when galaxies built up the majority of their stellar mass. Using the MeerKAT DEEP2 field---the deepest $\sim$1.4\,GHz radio image yet taken---\cite{matthews21a} measured brightness-weighted radio source counts $S^2n(S)$ down to $0.25\,\mu$Jy. As shown in \cite{condon18}, the brightness-weighted source counts $S^2n(S)$ of either SFGs, AGNs, or both are proportional to the luminosity function of the respective population integrated over all redshift:
\begin{equation}
    S^2n(S) = \frac{D_\mathrm{H_0}}{2\pi\ln(10)}\int_0^{\infty}u_\mathrm{dex}(L_\nu|z)\left[\frac{(1+z)^{\alpha-1}}{E(z)}\right]dz,
\end{equation}
where $D_{H_0}\equiv c/H_0$ is the Hubble distance, $\alpha\equiv+d\ln\,S / d\,\ln\,\nu$ is the spectral index, $L_\nu = 4\pi D_\mathrm{C}^2(1+z)^{1-\alpha}S$ is the luminosity per unit frequency, $D_\mathrm{C}$ is the comoving distance, $E(z) = [\Omega_\mathrm{m}(1+z)^3 + \Omega_\Lambda + \Omega_\mathrm{r}(1+z)^4]^{1/2}$, and the energy density function is
\begin{equation}
    u_\mathrm{dex}(L_\nu|z) \equiv L_\nu\rho_\mathrm{dex}(L_\nu|z)=L_\nu g(z) \rho_\mathrm{dex}\left[\frac{L_\nu}{f(z)}|0\right],
\end{equation}
where $g(z)$ and $f(z)$ represent density and luminosity evolution, respectively. If the source counts are computed for SFGs and AGNs separately using independently derived evolutionary functions, the total source counts---the quantity typically observed---are simply the sum of the source counts from the two populations.

\cite{matthews21b} determined the evolutionary functions $f(z)$ and $g(z)$ such that the local luminosity function $\rho_\mathrm{dex}(L_\nu|0)$ evolved backwards matches the observed source counts. At any redshift $z$, the SFRD is proportional to the product $f(z)g(z)$ through the FIR/radio correlation. In this way, \cite{matthews21b} constrained the star formation history of the universe for a global population (i.e. there was no information on individual galaxies) by modeling the source counts down to $0.25\,\mu$Jy. The resulting SFRD evolution measured from the deep MeerKAT radio observations is not only stronger than 
 SFRD evolution based on UV/optical measurements, but also $\gtrsim$50\% stronger than combined UV--IR measurements across all redshifts---deepening the divide between longer and shorter wavelength pictures of the star formation history of the universe.

Upgrades to the Very Large Array and the advent of science observations with the MeerKAT telescope have unlocked the potential to use radio continuum as a star formation tracer to cosmologically significant redshifts. Recent catalogs of radio continuum sources have been used to probe the SFRD, and there is an emerging scatter amongst the measurements and their implications for the star formation history of the universe. While some radio studies find agreement with UV--IR measurements of SFRD$(z)$ \citep[e.g.][]{novak17,ocran20,vandervlugt22}, others have found an increase over the UV--IR measurements \citep[e.g.][]{leslie20,matthews21b,enia22,cochrane23}. These studies illustrate many of the difficulties in using radio continuum to trace star formation across redshift: stacking is usually necessary to reach the flux density sensitivity needed to detect star-forming galaxies (SFGs), relying on photometric redshifts to derive luminosity functions and galaxy characteristics, and observing at lower frequencies where the relationship between radio luminosity and SFR is less understood.

This work introduces a comprehensive multiwavelength follow-up campaign of the MeerKAT DEEP2 field. By surveying the same field whose source counts implied stronger SFRD evolution than UV--IR measurements we minimize systematics and robustly test the evolutionary models of \cite{matthews21b}. Precise redshifts from fitting a combination of low-resolution spectra and optical/NIR photometry of 3,839 galaxies with corresponding MeerKAT detections confirm the cosmic radio SFRD modeling from the source counts and its discrepancy with the UV--IR picture. The results presented here establish a rich framework for dust-unbiased tests of SFR evolution and its diagnostics down to faint, normal galaxies responsible for amassing most of the stellar mass in the universe.

This paper is organized as follows. Section \ref{sec:obs} details the multiwavelength, spectrophotometric data of the MeerKAT DEEP2 field. Redshifts are derived from a novel SED fitting technique described in Section \ref{sec:sed}. We calculate radio luminosity functions from $0.2<z<1.3$ and compare them with models in Section \ref{sec:radioevo}. We discuss the implications of confirming radio-based models of stronger SFRD evolution in Section \ref{sec:implications}.

Absolute quantities were calculated for the flat $\Lambda$CDM universe with $H_0=70\,\mathrm{km\,s^{-1}\,Mpc^{-1}}$ and $\Omega_\mathrm{m}=0.3$. Our spectral-index sign convention is $\alpha\equiv + d\ln\,S / d\ln\,\nu$.


\section{Multiwavelength observations of the MeerKAT DEEP2 field}\label{sec:obs}

The MeerKAT DEEP2 field was observed as part of commissioning the MeerKAT array \citep{mauch20}. The field covers a $\Theta=69.2'$ diameter half-power circle centered on J2000 $\alpha=04$:13:26.4, $\delta=-$80:00:00 with a $\theta_\mathrm{1/2}=7\farcs6$ circular Gaussian synthesized beam. The location of the field was optimized to avoid bright radio sources whose large flux densities---combined with systematic position and gain uncertainties of the telescope---limit the achievable dynamic range. About 160 hours of integration resulted in a final thermal noise of $\sigma_n = 0.56 \pm0.01\, \mu\mathrm{Jy\,beam}^{-1}$. The wideband DEEP2 image is the average of 14 narrow subband images weighted to maximize the signal-to-noise (S/N) of sources with spectral index $\alpha=-0.7$, resulting in an effective frequency of $\nu=1.266\,$GHz. The confusion distribution of faint sources has such a long tail that it is not well-described by its rms. Using the traditional definition of the confusion limit as the flux density at which there are $\gtrsim$25 beam solid angles per source, the ``rms'' confusion noise is $\sigma_\mathrm{c}\approx 2.6\,\mu\mathrm{Jy\,beam^{-1}}$.

\cite{matthews21a} describes the construction of the MeerKAT-DEEP2 radio source catalog of 17,350 components down to $10\,\mu$Jy. In brief, \cite{matthews21a} applied the Obit \citep{obit} task FndSou, which decomposes islands of contiguous pixels into circular Gaussian components. Most sub-mJy radio sources have angular diameters $\phi \ll 1''$ \citep{cotton18,murphy11}, so all sources were treated as unresolved point-sources. This approximation was supported through qualitative and quantitative comparisons of point-source-only simulated images with the observed MeerKAT-DEEP2 field. The radio component catalog extends down to $S_\mathrm{1.266\,GHz} = 10\,\mu$Jy and includes corrections to positions and flux-densities due to the presence of confusion. \cite{matthews21a} used the confusion amplitude distribution to constrain the counts of unresolved sources as faint as $S=0.25\,\mu$Jy. We refer the reader to \cite{matthews21a} for more details.
 
Here we present initial results from a suite of photometric and spectroscopic observations of the MeerKAT DEEP2 field from the optical through near-infrared. A detailed account of the multiwavelenth observations and a corresponding catalog will be available in Matthews et al. 2024 (in prep). Below is a brief summary of the multiwavelength data contributing to the initial results presented here.

\begin{figure*}
    \centering
    \includegraphics[trim={8mm 3mm 2cm 1cm}, clip, width=0.95\textwidth]{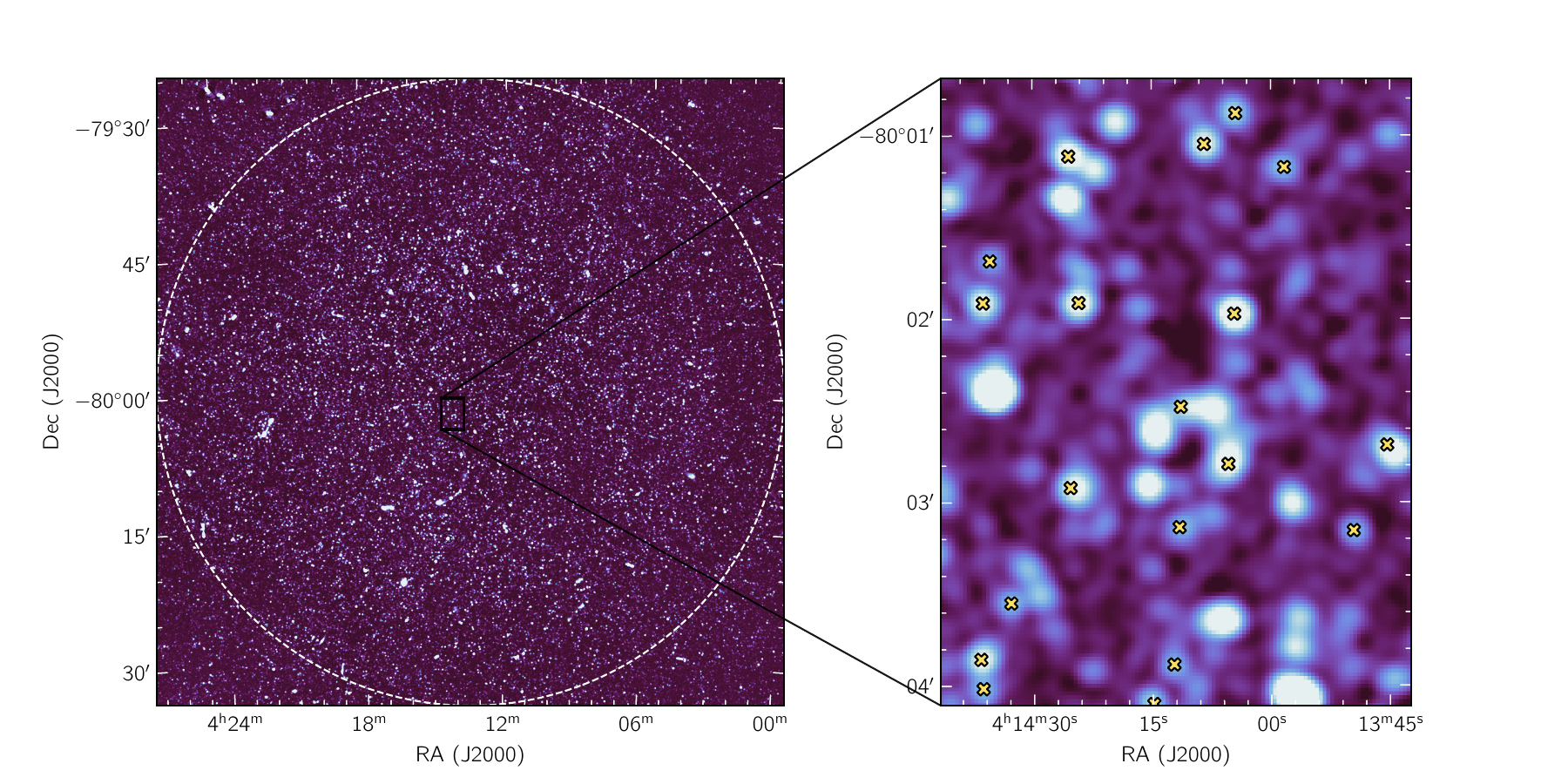}
    \caption{The 1.266\,GHz MeerKAT DEEP2 field is shown in the left panel---uncorrected for primary beam attenuation---with the half-power circle of the primary beam marked by the white dashed circle. Positions of the radio galaxies targeted for optical prism spectra are shown as yellow crosses. The right panel shows an enlarged version of a central $\sim3\farcm4\times\sim2\farcm6$ region. The initial data release of 3,839 redshifts represents a small fraction of the total radio sources.
    \label{fig:largedeep2}}
\end{figure*}


\subsection{Optical Prism Spectroscopy}

Obtaining redshifts for thousands of galaxies typically implies a photometric redshift approach; model or empirical template SEDs are fitted to fluxes in broadband filters to determine the physical properties of galaxies. With an effective spectral resolution of $R<10$, current and near-future large imaging surveys achieve uncertainties of at best $\sigma_z\gtrsim0.025(1+z)$ \citep[see][for a review of the topic]{newman22}.

A spectrophotometric approach adds low-resolution ($R\sim30$) optical prism spectra---which can be obtained for thousands of galaxies in one observation---to the fluxes in broadband filters. The continuous wavelength coverage in the optical with broadband filters extending into the NIR enable redshift uncertainties $\sigma_z\leq0.01(1+z)$, a method developed and implemented successfully in \cite{patel09, coil11, kelson14}. We adopt this spectrophotometric approach needed to accurately characterize the $\sim$17,000 galaxies in our radio-selected sample.

The Uniform Dispersion Prism (UDP) on the Inamori Magellan Areal Camera and Spectrograph \citep[IMACS;][]{imacs} at Las Campanas Observatory produces low-resolution spectra ($R = \lambda/\Delta\lambda \sim 30$) from 4000--9500 \AA. Because these spectra span only 150 pixels on the detector, between 1000--2000 objects can be simultaneously observed. 

Fourteen pointings of the IMACS f/2 camera with the UDP covered all of the MeerKAT DEEP2 half-power circle (the dashed line in Figure \ref{fig:largedeep2}). Each of the fourteen masks contained between 1,100 and 1,500 objects, with several objects duplicated between masks of overlapping regions. We demanded that each radio target have a counterpart in the 3.6$\,\mu$m \textit{Spitzer} image (described in Section \ref{sec:spitzer}) to register the radio positions to optical/IR astrometry. In total, 11,671 unique radio sources were observed and are identified with yellow crosses in Figure \ref{fig:largedeep2}. The median exposure time of the final sample is 12,560 seconds with a minimum exposure time of 4,000 seconds and 95\% of the objects having exposure times greater than 7,200 seconds. 

To guarantee that most of the slits placed on radio sources returned optical spectra, the initial round of objects (the first 7 of 14 masks) were chosen from the subset of the \cite{matthews21a} MeerKAT DEEP2 radio source catalog that had Spitzer cross-identifications within $1''$. This criterion applied for all of the first 7 of 14 masks and 2/3 of the objects on the last 7 of 14 masks. For the remaining 1/3 of objects on the last 7 of 14 masks, we relaxed the criterion to include a random selection from a preliminary ``deblended'' radio catalog made using the XID+ algorithm by \cite{hurley17}. Since these objects make up $\leq15\%$ of the targets (and an even smaller fraction of the reduced spectra) details of the deblending and cross-identifications will be described in a future work.
 
Observations were taken across 8.5 nights spanning February 2022 through January 2023. Typical exposure times for each mask ranged between 160 and 210 minutes.

\subsection{Ground-based Optical and Near-Infrared Photometry}

As part of programs 2022A-771331 and 2022B-132648, the MeerKAT DEEP2 field was observed by the Dark Energy Camera (DECam) in filters $ugrizY$. Observations spanned 3.5 nights from February 2022 through December 2022. We used the DECam Community Pipeline stacked image product for photometric measurements \citep{valdes14}. 

Poor atmospheric conditions persisted through most of the observations and limited the resulting seeing FWHM to $\gtrsim1\farcs2$ in all but $i$ band (where it reached an average seeing of $\sim0\farcs9$). This significantly degraded the achieved depth of the images, resulting in $\sim$5$\sigma$ limiting AB magnitudes of 24.5, 25, 24.75, 24.5, 24.0, and 23.0 for $ugrizY$, respectively. 

FourStar is a near-infrared (1--2.5\,$\mu$m) camera on the \textit{Magellan} Baade telescope at Las Campanas Observatory \citep{persson13}. It has a field-of-view of $10\farcm8\times10\farcm8$ composed of four $2048\times2048$ pixel detectors with a 19$''$ gap in between them. Covering the $\sim$1.1\,deg$^2$ MeerKAT DEEP2 field required a tiling pattern of 36 FourStar footprints. 

Observations were taken in the $J$-band filter (1.1--1.4\,$\mu$m) over 2.5 nights from February 2022 through December 2023. The average seeing over multiple nights of observation was $1\farcs1$. Each pointing was dithered 9 times in a square pattern around the central position with dither offsets of 26$''$ to cover the 19$''$ gap between the detectors. Data reduction and image processing was done using the custom pipeline FourCLift developed and described in \cite{kelson14}.

\subsection{Warm-Spitzer 3.6\,$\mu$m and 4.5\,$\mu$m imaging}\label{sec:spitzer}

The observations were carried out in twelve distinct epochs (Astronomical Observation Requests or AORs) from 8 July 2019 through 11 December 2019 and totaled 69.4 hours (\emph{Spitzer} PID:14246). Each AOR consisted of a 14$\times$14 mosaic of 100\,s frames, using a single point from the cycling dither pattern to break up the mosaic grid pattern. We included small ($\sim10''$) offsets in the central position, and, because the field is close to the southern continuous viewing zone (CVZ), there was a spread in position angle of the individual mosaics. All together, this ensured full coverage of the circular field containing the FWHM of the MeerKAT primary beam sensitivity (although, sources can be detected beyond this). The reduction of the IRAC photometry was completed in the standard way using the MOPEX data reduction package \citep{mopex}.

\subsection{Photometric Measurements}\label{sec:phot}

All astrometry was registered to Gaia DR3. Our deepest optical data were taken in the DECam $i$ filter, so we used this image to detect optical counterparts in the MeerKAT DEEP2 field. We used Source Extractor \citep{bertin96} to calculate and subtract the background and subsequently detect sources using the default parameters. We ran aperture photometry on the resulting $\sim10^6$ detections with a $4''$ diameter aperture to ensure all source flux of galaxies at intermediate redshifts and beyond is encompassed. The magnitude zeropoints in $grizYJ$ were simultaneously fit to match the synthesized stellar locus according to the algorithms in \cite{kelly14}. The $u$-band zeropoint was then calibrated using the ``blue tip'' of halo stars as described in \cite{liang23}.

\begin{figure*}
    \centering
    \includegraphics[width=0.95\textwidth]{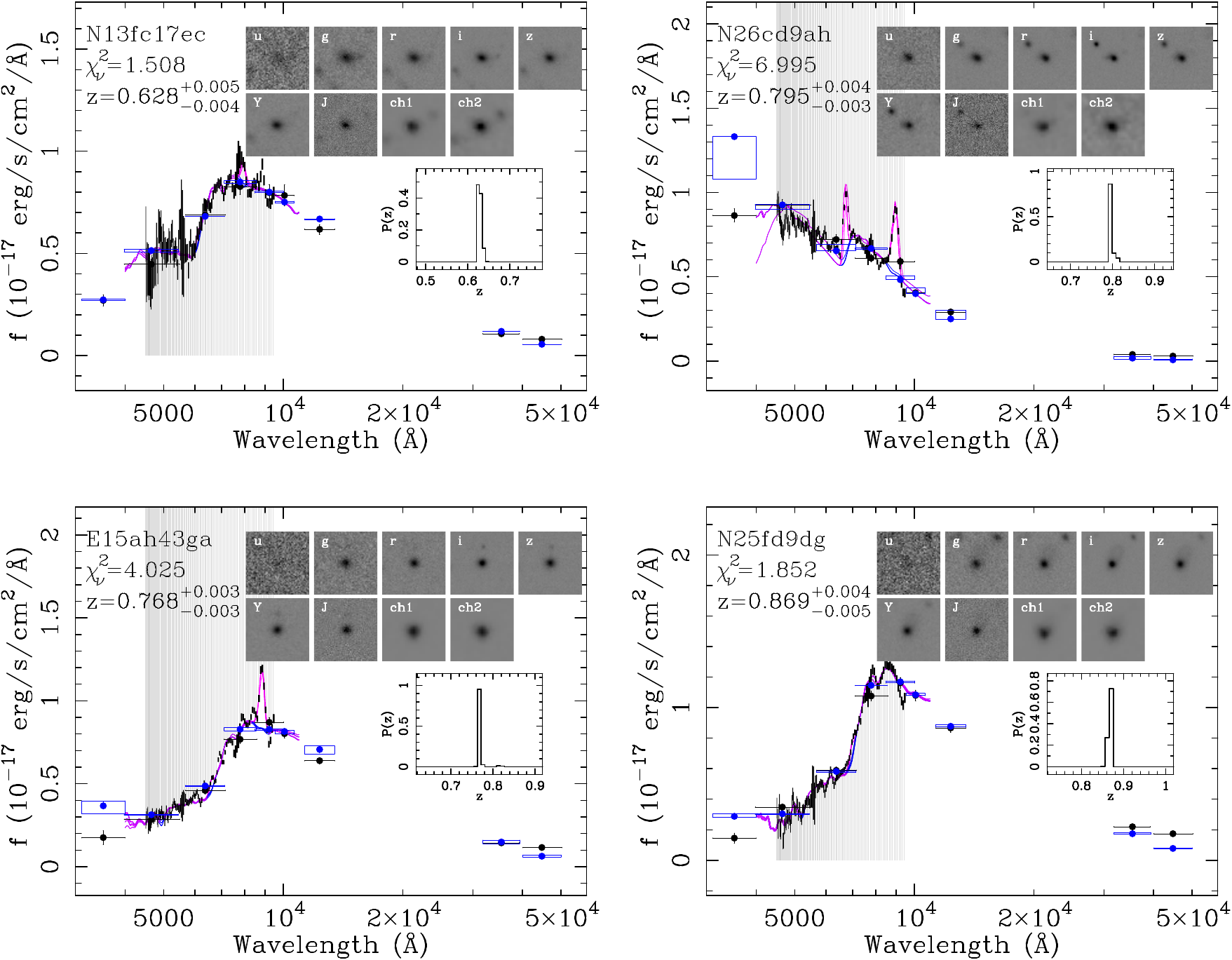}
    \caption{Four examples of radio galaxy SEDs are shown with their multiwavelength photometry (as $14''\times14''$ cutouts) and redshift probability distribution. Observed spectra and photometric flux values are shown in black. Our best-fitting model spectra and predicted photometric flux values are shown as the magenta line and blue points, respectively.}
    \label{fig:seds}
\end{figure*}

\section{An eigenvector approach to SED modeling}\label{sec:sed}

Constraining galaxy evolution necessitates accurate physical parameters from SED fitting. The accuracy and complexity can come at the expense of longer computation times, particularly for large samples of thousands of galaxies. We developed a new algorithm for SED fitting based upon the fact that SEDs can be constructed from a linear combination of basis functions. The details of this method will be fully described in Kelson et al. (in prep). In summary, we choose to abandon the idea that each basis function has a physical meaning and rather utilize the basis functions as eigenvectors that can be combined in various magnitudes to match a wide variety of galaxy SEDs. 

\subsection{Construction of Eigenvectors}

We first establish a grid of redshift, metallicity, and reddening values. There are 501 redshifts at intervals of $\Delta\log\,z=0.005$ between $-2 \leq \log\,z \leq 0.5$, six  metallicity values spaced at 0.3 dex intervals from -1.2 to 0.3, and eleven increments of $\log\,A_\mathrm{V}$ spaced at 0.25 dex intervals between $-2\leq\log\,A_\mathrm{V}\leq0.5$. Each point in the redshift-metallicity-reddening grid has 100,000 star-formation histories generated stochastically via the method introduced in \cite{kelson18}. We constructed our model SEDs using the Flexible Stellar Population Synthesis \citep[FSPS;][]{conroy09,conroy10} models with the Chabrier IMF \citep{chabrier03}. and adopted the reddening law of \cite{cardelli89}, which gave the best results (on average). In future iterations of this novel fitting method, we will be incorporating finer increments in metallicity and reddening, and exploring variation in the reddening law.

Singular value decomposition (SVD) of the resulting galaxy SEDs produces a plethora of basis functions, but many fall below observational noise and are discarded (for example, the signal-to-noise of the present sample requires three to six). To capture the diversty of stellar populations to a part in a thousand requires 9 eigenvectors. The spectroscopy---and photometry---here, like most survey data, are not accurate to a part in a thousand with a full accounting of flux calibration and photometric uncertainties, allowing us to restrict the fits to fewer basis functions without any loss of information. 



\subsection{Data Fitting}


Every observed galaxy SED is fit using a linear superposition of up to $N \leq 9$ eigenvectors plus five nonnegative emission line vectors. The maximum number of eigenvectors $N$ per galaxy is determined when the reduced $\chi^2$ is no longer improved by adding more eigenvectors. The superposition of eigenvectors accurately recreates the full diversity of galaxy stellar populations and we include the following five (non-negative, Gaussian) emission lines in order to capture the full diversity of galaxy SEDs (see, e.g., Fig. \ref{fig:seds}): H$\alpha$, [\ion{N}{2}], [\ion{O}{3}] (a superposition of 5007\AA\ and 4959\AA\ in the ratio 3:1), H$\beta$, the [\ion{O}{2}] doublet at 3727\AA\ (treated as two equal-magnitude Gaussians separated by 2.7\,\AA\ in the restframe), and Mg 2959\AA.

The final number of source SEDs fit using our novel eigenvector approach is 9,416, down from the original 11,671 due to criteria on deep photometry in multiple bands. Specifically, detection in [3.6], $J$--, and $i$--band were required as well as measurements in $r$, $i$, and $z$ brighter than 25, 24.5, and 24 AB mag, respectively, so as to ensure that the individual flux-calibration polynomials could be well constrained. After SED-fitting, we imposed additional constraints to be conservative in this first analysis of the data. These constraints included: (a) the 25th and 75th percentile of the S/N spectrum be greater than 0.25 and 1.0, respectively---this demands that the spectrum is not overly corrupted by random data issues that can corrupt sky subtraction (e.g., slit edge artifacts or scattered light from neighboring alignment star); (b) the reduced $\chi^2$ at a given S/N be less than 2 standard deviations from the median at that S/N ratio---to keep objects within the nominal data quality distribution; (c) objects do not lie on or near the stellar locus, as some stars were not excluded from the catalog prior to observation; (d) objects have narrow $P(z)$ distributions (we consider objects with overly broad $P(z)$ to have had too low S/N to be considered information of value; and (e) that the reduced $\chi^2 < 20$, which not only removes bright stars, but also high-redshift quasars. After these conservative cuts, our final sample is 3,839 galaxies. Four sources are shown for context in Figure \ref{fig:seds}. We are encouraged by the strong combination of efficiency and accuracy using this method. 

\subsection{Redshift Uncertainties}


The location of the MeerKAT-DEEP2 field was chosen for its potential to achieve unparalleled sensitivity---regardless of the availability of multiwavelength data. Unfortunately, being a field not previously covered by wide-field legacy surveys means that there is no extensive data set with high-resolution optical spectra or redshifts. Instead, we derive uncertainties using the large number of objects with repeat measurements.

We re-ran the SED fitting using UDP spectra of individual mask observations rather than using the combined spectra for objects with repeated measurements. There are 1,470 objects with repeated observations. We examined the biweight scatter in duplicate redshift measurements in redshift slices and find that the redshift errors are $\lesssim2\%$ in $(1+z)$ for all observations. 
The vast majority of combined spectra of sources used in the subsequent analysis have S/N $>10$. If we limit the duplicate measurements to this regime, the median redshift error is $\sigma_z \sim 0.006(1+z)$. 

The tails of the distribution of the redshift differences between two observations encode information on the catastrophic failures. The total fraction of objects where the duplicate redshift is more than 10\% off in $(1+z)$ is 5\%. 
For duplicate observations with $\mathrm{S/N}>10$---representative of the deeper spectra when all the data have been combined---the catastrophic failure rate is $<4\%$ for galaxies with $0.4<z<1$. In our lowest and highest redshift slices it is $<8\%$---a result of the poor seeing, and resulting insufficient depth of the DECam imaging.

\section{Radio source evolution}\label{sec:radioevo}

Comparisons of the space densities of galaxies as functions of luminosity (i.e. luminosity functions) at different points in time trace cosmological evolution. Models correctly describing this evolution thus constrain the buildup of stellar mass over cosmic time. Below, we describe the completeness of our sample---in both radio (Section \ref{sec:radiocompleteness}) and optical (Section \ref{sec:opticalcompleteness})---and the corrections we made to overcome inevitable incompleteness. Finally, in Section \ref{sec:rlf} we recount our luminosity function derivation and discuss the effects (or lack thereof) due to cosmic variance (Section \ref{sec:cosmicvariance}).

\subsection{Radio Reliability and Completeness}\label{sec:radiocompleteness}

The sensitivity of the MeerKAT DEEP2 image is limited by point-source confusion (rms $\sigma_c\sim2.6\,\mu$Jy everywhere), not by noise (rms $\sigma_n=0.56\,\mu$Jy at the pointing center increasing to $1.12\,\mu$Jy at the primary beam half-power circle). Catalogs of individual sources fainter than $S \sim 10\,\mu\mathrm{Jy}\sim 4\sigma_c$ become increasingly incomplete and unreliable. In the case of MeerKAT DEEP2, the catalog was cut off at $S_\mathrm{1.266\,GHz} = 10\,\mu\mathrm{Jy}$ ($\sim$18$\sigma_n$) to ensure high completeness and reliability as determined through injecting sources into mock images \citep{matthews21a}. Because the limiting flux density of the radio source catalog is independent of the thermal noise, the catalog sensitivity and completeness is uniform across the image. This translates to exceedingly simple radio completeness corrections that depend on flux density alone.

Ten mock images of the MeerKAT DEEP2 field were created by \cite{matthews21a} and cataloged with the same source finding algorithm employed on the real data. The completeness of the real source catalog was determined by calculating the fraction of input sources that were recovered and cataloged by the source-finding algorithm. The catalog at the faint limit of $S_\mathrm{1.266\,GHz}=10\,\mu\mathrm{Jy}$ is 57\% complete and quickly jumps to $\gtrsim96\%$ complete by $S_\mathrm{1.266\,GHz} = 20\,\mu\mathrm{Jy}$. \

The radio sample reliability $0\leq R\leq1$ is less than one because some of the sources with measured $S_\mathrm{1.266\,GHz} \gtrsim10\,\mu$Jy are actually fainter than the sample limit and should not be in the sample. This is true for any radio catalog, but is of particular concern for confusion-limited images. Using the ten mock images presented in \cite{matthews21a}, we estimate the reliability in 0.2 dex wide bins of $\log\,S$ centered on $\log\,S_\mathrm{1.266\,GHz} = -4.9,-4.7,-4.5,\cdots$. For cataloged sources in each $\log\,S$ bin, we calculate the fraction of the cataloged sources whose input source flux density is truly greater than $9.45\,\mu$Jy (thereby excluding objects whose flux was enhanced by more than the rms thermal noise). We add an additional constraint that for each cataloged source in a $\log\,S$ bin, the input source flux density truly lies within the $\log\,S$ bin in question. In this way, we estimate not only the reliability that a source is truly above the sample limit, but also the reliability that sources truly belong in their resident flux density bin (i.e. accounting for situations when the true flux density would place the source in the next highest flux density bin, but the cataloged flux was underestimated (e.g. due to negative sidelobes). The radio source completeness and reliability is shown in Figure \ref{fig:completeness}. 

\begin{figure}
    \centering
    \includegraphics[width=0.47\textwidth]{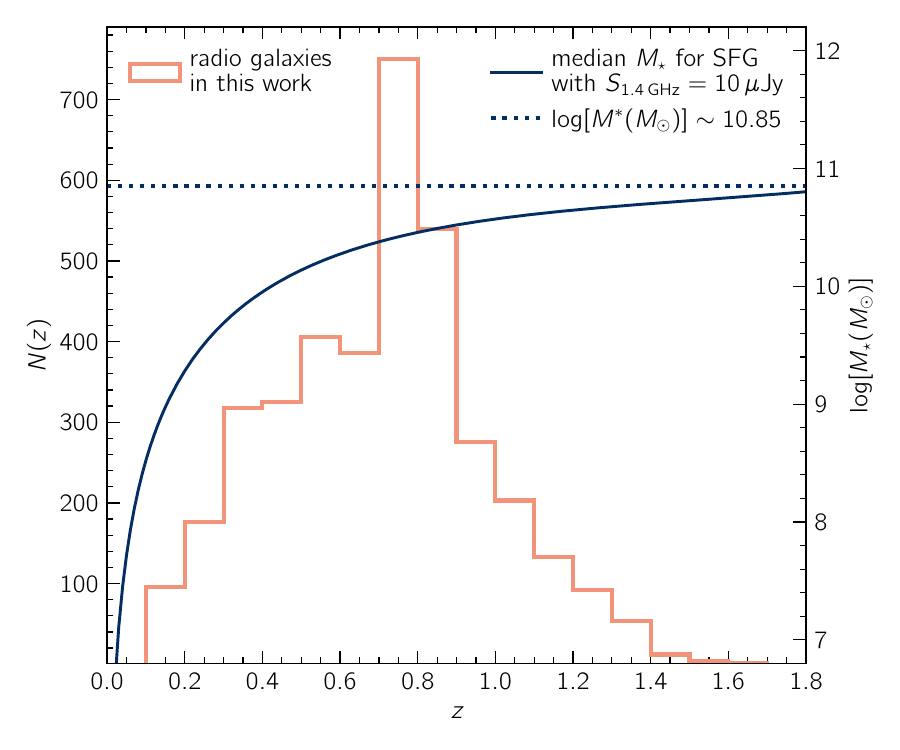}
    \caption{The redshift distribution of the final 3,839 MeerKAT DEEP2 sources presented in this work (orange curve). The median stellar mass of a star-forming galaxy with $S_\mathrm{1.4\,GHz}=10\,\mu$Jy as determined by the star-forming ``main sequence'' \citep[taken from][]{whitaker12} is shown as a solid blue line.}
    \label{fig:zdist}
\end{figure}

The radio catalog limit of $S_\mathrm{1.266\,GHz}=10\,\mu$Jy sets the minimum luminosity---and therefore the minimum SFR---this survey is sensitive to as a function of redshift. Assuming our radio sources are drawn from the population of normal star-forming galaxies and that SFR is correlated with stellar mass, we can use the parameterization of \cite{whitaker12} to find the stellar mass that corresponds to our survey flux limit as a function of redshift (shown in Figure \ref{fig:zdist}). We assume the characteristic mass of the galaxy stellar mass function is $\log[M^*\,(M_\odot)]\sim10.85$, the middle of the range $\log[M^*\,(M_\odot)]\sim10.7-11$ reported by \cite{weaver23} for the COSMOS2020 sample. At a $z=1.3$, our radio flux density limit corresponds to a galaxy with stellar mass $\sim 0.63\,M^*$, meaning our completeness of normal galaxies drawn from a representative sample approaches $\sim50\%$ at $z=1.3$. Galaxies below this characteristic mass threshold do make it into the sample, but only if they have abnormally higher SFRs compared to the locus defined in \cite{whitaker12}. By redshifts $z\sim1.3$, galaxies enter the sample in ways that are not representative of the full distribution of galaxies that fully contribute to the SFRD.

\subsection{Optical Completeness}\label{sec:opticalcompleteness}

The poor seeing prevented our DECam observations from reaching the intended $i=25$ magnitude limit. Nonetheless, because the seeing and depth of the $i$-band data are superior to the other bands, we use it to define the positions of galaxies associated with the radio detections for the photometry behind the SEDs. Cross-identifying optical counterparts in the $i$ band image yields 14,679 matches within $2\farcs5$ (smaller than half the $7\farcs6$ FWHM of the MeerKAT-DEEP2 beam) of the 17,150 radio galaxies. The Poisson probability that one or more unrelated radio sources lie within $2\farcs5$ of an $i$ band source is
\begin{equation}
 P(\geq1) = 1-P(0)=1-\exp(-\pi \rho r_s^2),   
\end{equation}
where $\rho$ is the sky density of sources with flux density brighter than a given value ($N(>\,S)$) and $r_s$ is the cross-matching radius. For the sky density of sources brighter than $S_\mathrm{1.266}=10\,\mu$Jy, $P(\geq1)\approx0.028$. The rate of false matches is not expected to impact the results.

\begin{figure}
    \centering
    \includegraphics[width=0.45\textwidth]{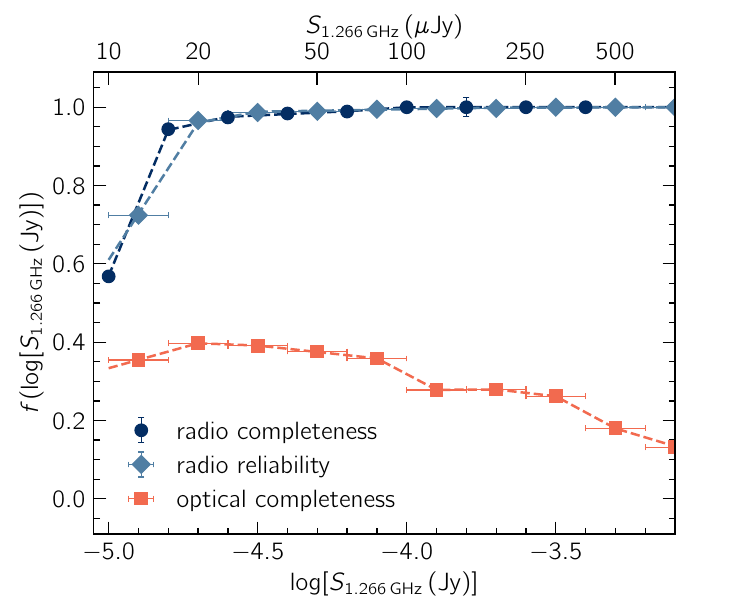}
    \caption{Radio completeness $C_\mathrm{r}(S)$, radio reliability $R_\mathrm{r}(S)$, and optical completeness $C_\mathrm{opt}(S)/f_{z<1.3}(S)$ as a function of flux density. The optical completeness includes the correction $f_{z<1.3}(S)$ accounting for the fraction of objects at $S$ that are expected to lie at $z<1.3$, for the purpose of avoiding over-correcting for incompleteness and inflating the resulting luminosity functions.}
    \label{fig:completeness}
\end{figure}

The fraction of these 14,679 galaxies with photometric measurements and successful SED fitting represents the optical completeness of the radio sample. 3,839 galaxies had both IMACS spectra free of data artifacts, high enough S/N, sufficient spectroscopic and photometric data quality and fidelity, and photometric measurements in a majority of the filters to fit an SED. We partitioned the 14,679 galaxies into $\log S$ bins of width 0.2 from $\log[S\,(\mathrm{Jy})]=-5$ to $\log[S\,(\mathrm{Jy})]=-2$. For each bin, the completeness is defined as the ratio of the number of galaxies with a successful SED fit to the total number within each bin. The resulting optical completeness curve is shown in Figure \ref{fig:completeness}. We find a slight decreasing trend of optical completeness with increasing radio flux density, ranging from $\sim$40\% complete at the lower limit $\log[S\,(\mathrm{Jy})]=-5$ to $\sim$20\% at $\log[S\,(\mathrm{Jy})]=-3.1$. Comparing the optical completeness with $i$-band magnitude, we are most complete ($\sim$44\%) at a $i=21$ and the completeness falls near-linearly to $\sim$5\% at $i=24$. We limit our luminosity function calculations to $i=23.5$, the point at which the completeness falls to $12.5\%$, just over a quarter of its peak value.

\subsection{Radio luminosity functions}\label{sec:rlf}

We calculate luminosity functions in several contiguous redshift slices using the 1/$V_\mathrm{max}$ method \citep{schmidt68}. We define redshift bins to be wide enough such that the Poisson counting errors in that redshift slice are at most $\sim5\%$ ($N\approx350$), but narrow enough to mitigate the impact of evolving stellar populations. If redshift slices are too wide, some galaxies that may have sufficient photometry at the front end of a slice would have had photometric properties---owing to both mass growth and stellar evolution---at the back end of the redshift slice that leave them excluded by the DECam imaging depth and complicate the estimation of $V_\mathrm{max}$. Our sample has a median redshift error $\sigma_z\lesssim 0.01(1+z)$, which means the average uncertainties on $z$ are at most $\sim17.5$\% (for redshift bin $0.7<z<0.8$) of the width of the redshift bins, meaning objects are unlikely to scatter into neighboring redshift slices at a level affecting the luminosity function or SFRD measurement. 

The maximum comoving volume in which a galaxy can be observed $V_\mathrm{max} = V(z_\mathrm{max}) - V(z_\mathrm{min})$ is the volume of the redshift bin over which the source is observable. It is bounded on one side by $z_\mathrm{min}$, the lower bound of the redshift bin containing the galaxy. On the high end, the maximum redshift $z_\mathrm{max}$ is the minimum of the following: (1) the redshift at which the galaxy falls out of the radio sample ($S_\mathrm{1.266\,GHz} < 10\,\mu$Jy), (2) the redshift at which the galaxy falls out of the optical sample ($i > 23.5$---and assuming a galaxy's color does not evolve over the time span of its respective redshift  slice), or (3) the maximum redshift of the redshift slice containing the galaxy. 

The spectral luminosity $L_\mathrm{1.4\,GHz}$ at redshift $z$ depends on the observed flux density of the source and the spectral index of the source population
\begin{equation}
    L_\mathrm{1.4\,GHz}(z) = 4\pi D_L^2 (1+z)^{-\alpha-1}\left(\frac{1.4}{1.266}\right)^\alpha S_\mathrm{1.266\,GHz},
\end{equation}
where we assume the standard $\alpha = -0.7$ for radio populations at $\sim$1.4\,GHz. 

Unlike in the radio regime, the SED of a galaxy in the optical does not follow a power law and depends sensitively on the K-correction. 
\begin{equation}
    m_i(z) = M_i + \mathrm{DM}(z) + K_i(z),
\end{equation}
where $m_i$ is the apparent magnitude of the source at redshift $z$, $M_i$ is the $i$ band absolute magnitude, DM$(z)$ is the distance modulus, and $K_i(z)$ is the K-correction at $i$ band. We modeled the K-correction using the \texttt{kcorrect} software by \cite{kcorrect}. 
We assume the galaxy SED---the best-fit spectral template from \texttt{kcorrect}---does not change within its redshift bin and calculate the maximum redshift for which $m_i \leq 23.5$.

For each luminosity bin, the space density of radio sources is calculated by the following:
\begin{equation}\label{eq:vmax}
\rho_\mathrm{dex}(L,z) = \frac{1}{\Delta \log\,L}\sum_{i=1}^{N}C_i\left[V^{-1}_{\mathrm{max},i}\right],
\end{equation}
where $\Delta \log\,L$ is the width of the luminosity bin, $V_{\mathrm{max},i}$ is the maximum comoving volume over which the $i$th galaxy could be observed,
\begin{equation}
    V_\mathrm{max,i}= \frac{\Omega}{4\pi}[V(z_\mathrm{max,i}) - V(z_\mathrm{min,i})],
\end{equation}
where $\Omega$ is the survey area in steradians, and $C_i$ is the completeness correction factor (equal to 1/completeness) for the $i$th galaxy. Luminosity bins are of width 0.2 dex centered on $\log[L_\mathrm{1.4\,GHz}(\mathrm{W\,Hz^{-1}})]= 21.1, 21.3, \ldots, 25.3$. 

\begin{figure}
    \centering
    \includegraphics[trim={0mm 0mm 0cm 0cm}, clip, width=0.435\textwidth]{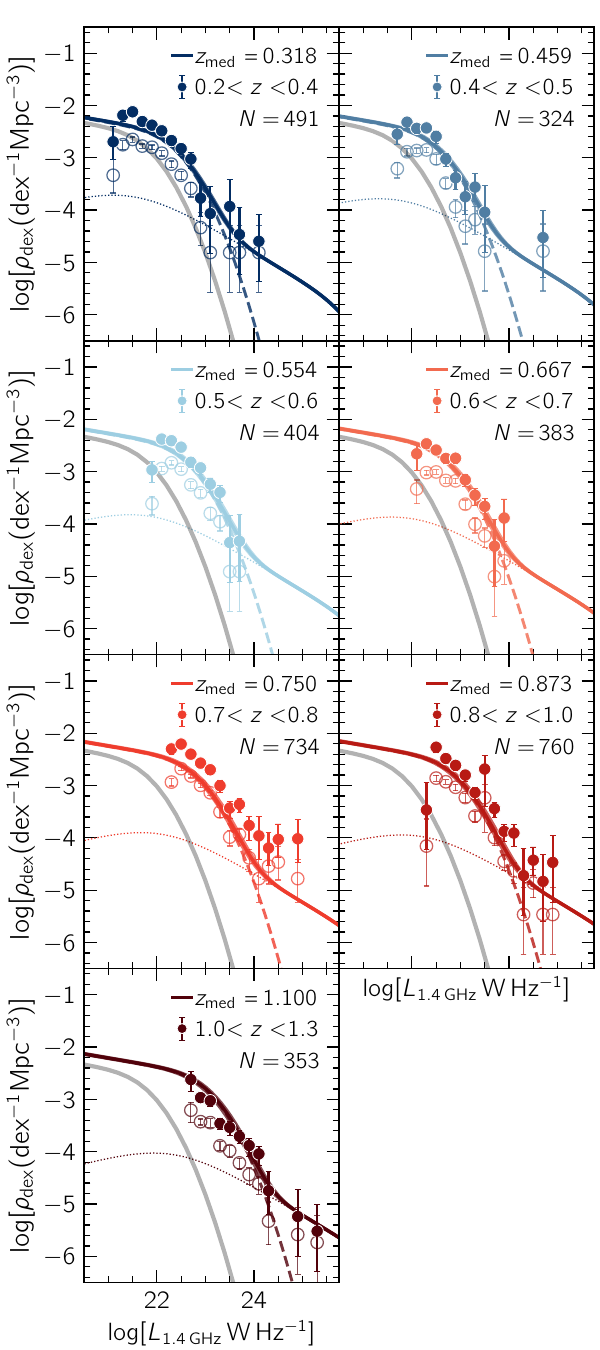}
    \caption{
    Observed number densities of radio sources (filled circles)---corrected for both radio and optical incompleteness---as a function of spectral radio luminosity. Open circles represent the number densities of the same sample without completeness corrections. Error bars reflect $1\sigma$ uncertainties. Solid colored line: radio luminosity functions predicted by \cite{matthews21b} for all galaxies (star-forming and AGN) at the median redshift of the observed galaxies in each redshift slice. The predicted radio luminosity functions of only SFGs or only AGNs at the median redshift are shown as the dashed and dotted lines, respectively. Thick gray line: local radio luminosity function of star-forming galaxies from \cite{condon19}.}
    \label{fig:rlf}
\end{figure}

The scenario in which our sample is 100\% complete---where we have measured redshifts for every radio source at $z<1.3$ and that all other sources are at $z>1.3$---is illustrated by the open circles in Figure \ref{fig:rlf}. 
Assuming all unmeasured objects are at $z<1.3$ overestimates the incompleteness and leads to measured luminosity functions far above the predicted luminosity functions calculated using the radio evolution models of \cite{matthews21b}. Here, we presume completeness corrections assuming that the true fraction of radio sources at $z<1.3$ is that implied by the modeling from \cite{matthews21b}. As a function of flux density, the fraction of all radio sources with $z<1.3$ ranges from 52\% at $\log[S\,(\mathrm{Jy})]=-5$, rises to a peak of 81\% at $\log[S\,(\mathrm{Jy})]=-3.5$, and plateaus around $\sim$70\% for sources with flux densities approaching 100\,mJy. The completeness correction factor $c_i$ for the $i$th galaxy includes a term for the reliability of the radio catalog as a function of flux density $R_\mathrm{r}(S)$, for the completeness of the radio survey as a function of flux density $C_\mathrm{r}(S)$, for the optical/redshift completeness of the radio sample as a function of flux density $C_\mathrm{opt}(S)$, and for the fraction of objects at flux density $S$ that lie within $z<1.3$ by the evolutionary models of \cite{matthews21b}. The final completeness correction factor for the $i$th galaxy is as follows:
\begin{equation}
    C_i = \frac{R_\mathrm{r}(S_i)\,f_{z<1.3}(S_i)}{C_\mathrm{r}(S_i)\,C_\mathrm{opt}(S_i)}.
\end{equation}

The errors are the quadrature sum of the expected fractional cosmic variance and rms Poisson counting errors for independent galaxies. If the number of galaxies in a luminosity bin is small ($N<5$), the counting errors are taken from the 84\% confidence limits tabulated in \cite{gehrels86}. The uncertainty caused by cosmic variance is described below in Section \ref{sec:cosmicvariance}. 

\subsection{Cosmic Variance}\label{sec:cosmicvariance}


We estimate cosmic variance using three independent methods: with the Cosmic Variance Calculator from \cite{trenti08}, from the radio sky simulations of \cite{bonaldi19}, and following the results of \cite{driver10}. \cite{trenti08} model the cosmic variance using a combination of analytic estimates via the two-point correlation function of dark-matter halos in an extended Press-Schechter formalism \citep{press74}, and numerically using synthetic catalogs from $N$-body simulations of large-scale structure formation. 
To connect the cosmic variance of the radio source population to the predicted clustering of dark matter halos from \cite{trenti08}, we use the stellar-to-halo mass relation presented in \cite{leauthaud12} and estimate the average halo mass associated with the stellar-masses of the radio sources determined from the SED modeling, which range from $\langle\log M_*\rangle = 10$ to $\langle\log M_*\rangle = 10.5$ from the lowest to highest redshift slice.

With the estimated average halo masses corresponding to our radio sample, we use the Cosmic Variance Calculator to estimate the 1$\sigma$ fractional cosmic variance in each redshift bin \citep{trenti08}. For the survey area of the MeerKAT-DEEP2 field ($\sim$1\,deg$^2$), the fractional count error due to cosmic variance is consistently less than 0.16 across all redshift slices in our sample. Tweaking the Cosmic Variance Calculator parameters (e.g. halo filling factor) has little effect on the output cosmic variance values and therefore inconsequential effects on the space densities of radio galaxies determined in this work. 

We independently calculate the cosmic variance using the Tiered Radio Extragalactic Continuum Survey (T-RECS) presented in \cite{bonaldi19}. T-RECS models two main populations of radio sources: SFGs and AGN over the 150\,MHz to 20\,GHz range. From the 25\,deg$^2$ ``medium'' deep tier, we extract non-overlapping circular sample areas of $\sim$1\,deg$^2$---the size of the MeerKAT-DEEP2 field. This amounts to 16 independent sky areas. In each simulated sky area, we combine both SFGs and AGNs---as we are currently unable to distinguish between the two populations in our observations of the MeerKAT DEEP2 field---and calculate number counts in the same 1.4\,GHz luminosity bins used to construct the observed luminosity functions. We find that the cosmic variance is consistent with the values estimated by the Cosmic Variance Calculator through $z=0.7$ and falls below the Cosmic Variance Calculator estimations for $z>0.7$. For $z>0.7$, the cosmic variance determined by comparing the number counts in the 16 sky areas of the \cite{bonaldi19} simulations is $\lesssim10\%$ and contributes only 0.04 dex to the error budget.

Finally, we refer to the cosmic variance estimates of \cite{driver10}, who used the Sloan Digital Sky Survey (SDSS) to empirically develop formulae for estimating cosmic variance based on survey shape and volume. It is important to note that the estimations from \cite{driver10} were derived for a $M^*\,\pm1$\,mag population of galaxies. This assumption captures most---but not all---of the sample of radio sources we present here. For the median redshift in each slice, we calculate the comoving transverse distance of the MeerKAT DEEP2 field radius and the comoving radial depth for the width of that redshift slice. Using these values and Equation 4 of \cite{driver10}, we calculate the percent cosmic variance in each redshift slice. For bins of $\Delta z=0.1$\,dex, the cosmic variance ranges from 24--28\%---much higher than that predicted by the other two methods. Wanting to be conservative with our systematic uncertainties, we adopt Equation 4 of \cite{driver10} to estimate the cosmic variance in our SFRD measurements and increase the minimum width of our redshift bins to $\Delta z=0.2\,$dex for the SFRD calculation.

\subsection{Estimating radio-based SFRD($z$)}

Converting radio luminosities to SFRs deserves extensive care and a thorough investigation of possibly uncertainties (as will be outlined in Section \ref{sec:implications}). However, to guide such discussion and quantify the possible disagreement in the SFRD($z$) between radio and UV--IR studies, we present an initial conversion between the calculated radio luminosity functions and SFRD$(z)$.

The SFRD at any redshift $z$ is related to the total radio energy density produced by star-forming galaxies. In each redshift slice, we fit the projected luminosity-weighted luminosity functions (e.g. energy density functions) of SFGs and AGNs combined to the observed data. As an initial estimate, we assume the form of the luminosity functions modeled by \cite{matthews21b} and only allow a renormalization of the SFG energy density function with respect to both axes (i.e. a scale and a shift). 

To properly compare the observed SFRD in each redshift slice with that predicted by the \cite{matthews21b} evolutionary models, we adopt their prescription to convert radio luminosity to SFR. In short, we assume the relationship between SFR and integrated IR luminosity ($8<\lambda(\mu\mathrm{m})<1000$) of \cite{murphy11} with the average ratio between total IR luminosity and FIR luminosity ($42.5<\lambda(\mu\mathrm{m})<122.5$) measured by \cite{bell03}. The conversion from the total radio spectral power at $1.4\,$GHz at any time $t$ to SFRD $\psi(t)$ for a \cite{kroupa01} IMF is 
\begin{equation}\label{eq:sfrd}
    \begin{aligned}
    \left[\frac{\psi(t)}{M_\odot\,\mathrm{yr^{-1}\,Mpc^{-3}}}\right] = 7.39\times 10^{-37}\cdot 3.75\times 10^{12}\,\mathrm{Hz}\\
    \cdot\left(\frac{U_\mathrm{1.4\,GHz}(t)}{\mathrm{W\,Hz^{-1}\,Mpc^{-3}}}\right)\cdot10^{\langle q(L_\nu)\rangle},
\end{aligned}
\end{equation}
where $\langle q(L_\nu)\rangle$ is the ratio of the FIR and radio luminosities from \cite{matthews21b}:
\begin{align}
    \langle q \rangle &= 2.69-0.147[\log(L_\nu)-19.1] \ \mathrm{if \ \log(L_\nu)<22.5} \notag\\
    \langle q \rangle &= 2.19 \ \mathrm{if \ \log(L_\nu)}\geq22.5.
\end{align}

We calculate $U_\mathrm{1.4\,GHz}$ directly by summing $L_\mathrm{1.4\,GHz}/(C\,V_\mathrm{max})$---where $C$ is the same completeness correction factor defined in Section \ref{sec:rlf}---over the unbinned sample of galaxies within each redshift slice ($0.2<z<0.4$, $0.4<z<0.6$, $0.6<z<0.8$, $0.8<z<1.0$, and $1.0<z<1.3$). We restrict the summation to galaxies with $L_\mathrm{1.4\,GHz} \leq 24.25\,\mathrm{W\,Hz^{-1}}$, the luminosity corresponding to an SFR of $\sim1000\,M_\odot\,\mathrm{yr^{-1}}$. This upper limit is chosen to minimize AGN contamination.

\begin{figure}
    \centering
    \includegraphics[trim={6mm 5mm 14mm 6mm},clip,width=0.47\textwidth]{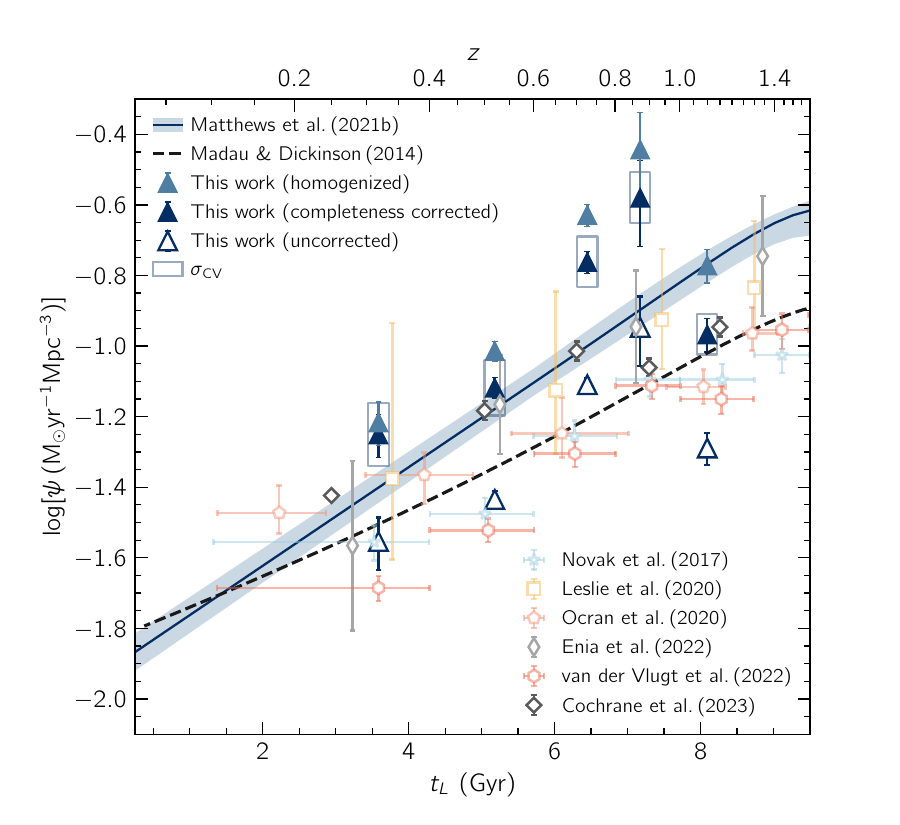}
    \caption{The SFRD from present day to a lookback time of $t_L=9.5$\,Gyr. A sample of UV--IR \citep{madau14,marchetti16} and recent radio \citep{leslie20,enia22,cochrane23} continuum SFRD measurements are shown for comparison. SFRD measurements from our sample of 3,839 galaxies in the MeerKAT DEEP2 field are shown as filled, thick triangles. Unfilled triangles show the SFRD measurements from the MeerKAT DEEP2 sample without any corrections for sample incompleteness. The uncertainties in each redshift slice due to cosmic variance---estimated from \cite{driver10}---are shown as the rectangular outlines around the dark blue triangles. All measurements have been converted to a Salpeter IMF \citep{salpeter55} for ease of comparison with the \cite{madau14} relation.}
    \label{fig:sfrd}
\end{figure}

Our calculation of $U_\mathrm{1.4\,GHz}$ is independent of any assumptions about the shape or evolution of the radio luminosity function, providing an essential test on the radio luminosity and density evolution models derived in \cite{matthews21b}. However, the imposed flux limit corresponds to a different minimum luminosity in each redshift slice and probes different fractions of the total energy output by star-forming galaxies. To homogenize our SFRD measurements, we use the models derived in \cite{matthews21b} to estimate the fraction of the total energy-density our survey is sensitive to for each redshift slice. We divide by this fraction---ranging from 70\% in the highest redshift slice to 92\% in the lowest---to correct for this incompleteness. For transparency, we include three versions of our radio-based SFRD measurement in Figure \ref{fig:sfrd} at the median redshift of each slice: (1) the raw measurements, uncorrected for incompleteness of any kind, (2) the measurements corrected for the completeness and reliability of the optical and radio surveys, and (3) the corrected measurements homogenized to probe an equivalent fraction of the total energy density.

There is excellent agreement between our independent SFRD measurements and both the SFRD predictions from the models in \cite{matthews21b} and SFRD measurements from other radio-based studies \citep{cochrane23, enia22,leslie20} in the range $0.2<z<0.7$. However, there is notable disagreement between the SFRD measurements presented here with other radio-based studies \citep[][]{novak17,ocran20,vandervlugt22}. Further work to (1) push to fainter flux densities (and therefore further down the faint end of the luminosity function) and (2) carefully separate SFGs and AGNs in the MeerKAT DEEP2 sample will expand the integration limits and place stronger constraints on the radio-SFRD($z$).




\section{Potential sources of discrepancy in the SFRD evolution and Implications}\label{sec:implications}



At the end of the previous section, we showed star formation rate densities---inferred from our luminosity functions---in Figure \ref{fig:sfrd}. These SFRDs---shown by the filled triangles---are compared to those from other radio studies \citep[e.g.][]{leslie20,enia22,cochrane23}, and to the \cite{madau14} model based on a large sample of UV--IR estimates. 
While this study advances our understanding of the radio-SFRD by coupling the the unprecedented sensitivity of the MeerKAT DEEP2 radio image with a large spectrophotometric sample, the tension with UV--IR measurements has been seen across multiple radio studies (such as those listed above). Given that the luminosity functions agreed with those predicted by \citet{matthews21b}, Figure \ref{fig:sfrd} also shows that our SFRD measurements agree with the SFRD evolution of the \citet{matthews21b} modeling.

The radio values for SFRD between $(0.2\leq z\leq 1.3)$ are 50\% to 100\% higher than canonical estimates from not only UV/optical, but also, seemingly, over combined UV--IR measurements. This discrepancy suggests more than a mild tension, but rather a fundamental disagreement, in one of the most important metrics for understanding the evolution of our universe. Further, there is emerging scatter amongst the radio-based SFRD measurements---with some in agreement with UV--IR measurements \citep[e.g.][]{novak17,ocran20,vandervlugt22} and others significantly discrepant in all or selected redshift regimes \citep[e.g.][]{cochrane23,matthews21b}. Here we discuss sources of systematic uncertainty on measurements of luminosity and space densities, quantify their impact on SFRD calculations, and explore possible reasons for disagreement between radio and UV--IR based SFRD$(z)$.

\subsection{Budget of systematic uncertainties}

A full budget of systematic uncertainties must include many potential issues, some of which have been mentioned already. In Section \ref{sec:cosmicvariance} the uncertainty due to cosmic variance was estimated using three independent methods, with resulting fractional cosmic variance uncertainties ranging from 23\% in the lowest-$z$ bin to 14\% in the highest-$z$ one, according to the most conservative of the three methods \citep{driver10}. Other potential sources of uncertainty include confusion effects, completeness correction factors, contamination by AGN, and the FIR/radio correlation parameter $q$. 

The volume surveyed here, and the modest $7\farcs6$ PSF, are both large enough to produce more than zero incidents of coincidence between galaxies and background radio sources along the line of sight. We generated mock MeerKAT images by populating the image with sources at redshifts and luminosities drawn from the model radio luminosity functions of \cite{matthews21b}. After a blind cataloging---without prior knowledge of injected source positions---of the image, we simulated the observed luminosity functions. Comparing the mock catalog with the positions and luminosities of input sources, we determined the occurance rate and effects of object coincidence per luminosity bin and as a function of redshift. At $z<0.7$ and $z>1$ the effect is negligible, while at $0.7<z<0.9$ the observed luminosity functions may be contaminated by line of sight superposition at a level that can, on average, skew the luminosities of sources by $+0.05$ dex.

There is an additional systematic uncertainty that arises from the modeling of spectroscopic incompleteness, as we have only securely measured redshifts for a fraction of the radio catalog. We estimate this uncertainty by summing the galaxy weights in our final catalog of $3,839$ galaxies and comparing to the size of what ought to be the parent catalog---the number of radio galaxies in the MeerKAT DEEP2 catalog multiplied by the fraction of those expected to have $z<1.3$. This latter term is model dependent and we adopt a flux-density dependent fraction from \citet{matthews21b}. We find the sum of galaxy weights (i.e. inverse completeness) to be within $\sim 10\%$ of the size of our parent catalog. As such, we estimate errors in the completeness correction factor to add $\sim$10\% uncertainty (0.04 dex) to our SFRD measurements.

The fraction of AGN in our radio sample will decrease with decreasing flux density. Further, since AGN components are omitted in our SED fitting method, AGNs exhibiting strong spectral features in the optical will result in poor SED fits and are therefore cut from our sample following our conservative data-quality criteria. Nonetheless, radio emission powered by AGNs is present in our sample at some level. After identifying AGNs with a wide variety of multiwavelength diagnostics, \cite{algera20} found that at $S_\mathrm{3\,GHz} = 30\,\mu$Jy the fraction of radio sources powered primarily by star-formation reaches $90\%$. At $1.266\,\mathrm{GHz}$, $S_\mathrm{3\,GHz} = 30\,\mu$Jy equals $\sim55\,\mu$Jy (assuming $\alpha=-0.7$). While 82\% of our sample lie below this flux density value, we conservatively assume 10\% of our radio emission is contributed by AGN. As such, AGN contamination add $\sim$0.04 dex to the total error budget on our SFRD measurements.

\subsection{Sources of disagreement in SFR conversions}

Disagreement between can arise for two reasons: problems in the measurements (UV, IR, and/or radio) themselves, or problems in the conversions from these measurements to SFRs. This work will focus on areas of improvement for radio-based determinations of the SFRD($z$). 


Particularly important to our work is the calibration of radio continuum as a tracer of SFR through the FIR/radio correlation. We have adopted the sub-linear relation calculated for SFGs in the local universe by \cite{matthews21b}. Had a luminosity-independent $\langle q\rangle=2.34\pm0.01$ been adopted from the seminal work of \cite{yun01}---using galaxies in the local universe---our SFRDs would increase further by $\sim$0.1 dex. The SFRD calculated by \cite{leslie20} (shown in Figure \ref{fig:sfrd}) also adopted a sub-linear IR/radio correlation \citep{molnar21}. \cite{molnar21} calculated the total-IR (TIR, $8<\lambda(\mu\mathrm{m})<1000$) to radio correlation, but---assuming the conversion between TIR and FIR by \cite{bell03}---this relationship yields consistent SFRD measurements when applied to our sample. 

\begin{figure}
    \centering
    \includegraphics[width=0.47\textwidth, trim={3mm 5mm 14mm 6mm}, clip]{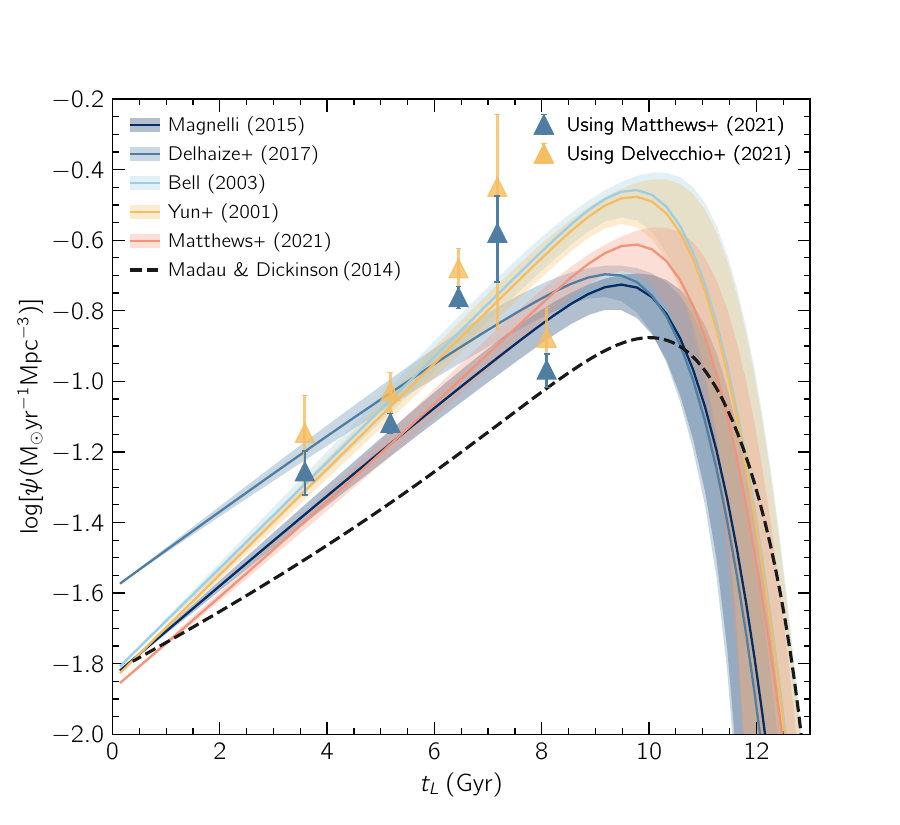}
    \caption{The SFRD$(z)$ is shown as a function of lookback time. Different prescriptions for the conversion from radio to FIR/TIR luminosity are explored as different colored curves. The blue points are the direct SFRD measurements from this work using the sub-linear FIR/radio correlation from \cite{matthews21b}. As a comparison, the gold points show the resulting SFRD measurements when the stellar-mass-dependent correlation found by \cite{delvecchio21} is applied to our sample.}
    \label{fig:firrc}
\end{figure}

Figure \ref{fig:firrc} demonstrates the range of resulting SFRD values that can occur when adopting various FIR/radio correlation prescriptions. Of the more modern---and robust---relations for $q$, the sub-linear relationship ($L_\mathrm{FIR}\propto L_\mathrm{1.4\,GHz}^{0.85}$) presented by \cite{matthews21b} and the stellar-mass dependent correlation found by \cite{delvecchio21} yield the largest SFRD values, but are consistent with the redshift-dependent $q$ relationships presented by \cite{magnelli15} and \cite{delhaize17} in $0.2<z<1.3$.

When added in quadrature the multiple sources of systematic error in our SFRD measurements yield $\sim 0.13$ dex. Given the combined random and systematic uncertainties, our measurements substantially agree with the model by \cite{matthews21b} of SFRD evolution (the redshift slices at $z \sim 0.7-1$ yield SFRD measurements $\sim 2-3\sigma$ high) and in significant disagreement with the SFRD characterized by ``canonical'' UV--IR estimators. We believe cosmic variance in the DEEP2 area is driving the jump in SFRD around $z\sim0.8$. Preliminary analysis of the spatial distribution of galaxies in each redshift slice suggest the presence of multiple galaxy groups or clusters around $z\sim 0.8$. A detailed quantification of this overdensity and the implications will be addressed in a subsequent paper.

While some sources of systematic errors cannot be mediated (e.g. cosmic variance), others can be improved. Further analysis of the MeerKAT DEEP2 field will explore the connection between radio luminosity and SFR across a wide variety of galaxy stellar masses and SFRs. Comparison of SFRs derived from SED fitting (some of which are supplemented by emission line fluxes measured directly from the spectra) with those by radio luminosity will illuminate areas of galaxy parameter space where these diagnostics disagree. Particular attention to resolving the cause behind these SFR disagreements will inevitably improve our understanding of dust-reddening evolution, cosmic-ray diffusion, and stellar mass buildup in star-forming galaxies.

\section{Summary}

We present initial results from the spectrophotometric multiwavelength follow-up of the MeerKAT DEEP2 field. Using a novel SED-fitting method on a combination of low-resolution spectra and optical/NIR photometry, we determined redshifts for 3,839 galaxies spanning $0.2<z<1.3$. This sample provides the first tests of critical assumptions---such as the shape of the luminosity function remaining constant with time---and predictions---such as the redshift distributions of radio sources---underlying the modeling of counts by \cite{matthews21b}. As can be seen from the luminosity functions in Figure \ref{fig:rlf}, the assumed shape of the luminosity function and predicted number density of radio sources was not grossly in error. We conclude with the following:
\begin{itemize}
    \item The luminosity functions measured from the collection of 3,839 individual radio galaxies agree remarkably well with the luminosity and density evolutionary models derived from only global (i.e. lacking any individual galaxy-level information) radio source counts \citep{matthews21b}.
    \item This agreement confirms the findings of \cite{matthews21b}---there is strong evidence for $\gtrsim50\%$ increased SFRD evolution at radio frequencies than what has been measured in the combined UV--IR. The source of this discrepancy remains unknown and is the focus of our continued work with these data.
    \item Radio measurements of the SFRD show increasing scatter, with some studies confirming enhanced SFRD evolution over UV--IR measurements and others in agreement with the UV--IR SFRD($z$) \citep[i.e.][]{madau14}. We hope that our continued multi-wavelength follow up of the MeerKAT-DEEP2 radio sources---spectroscopic and photometric---will provide further insights into the connections between the radio and UV--IR emission from star forming galaxies.
\end{itemize}




\acknowledgments
We thank the referee for a thorough and knowledgeable review of the paper which greatly improved the final work. AMM acknowledges support from a Vera Rubin Fellowship. The National Radio Astronomy Observatory is a facility of the National Science Foundation operated by Associated Universities, Inc.

\bibliographystyle{aasjournal}
\bibliography{paperv3}
\vfill\eject
\end{document}